# La$_{0.95}$Sr$_{0.05}$CoO$_3$: An efficient room-temperature thermoelectric oxide

J. Androulakis, P. Migiakis[a)], J. Giapintzakis[b),*]

*Institute of Electronic Structure and Laser, Foundation for Research and Technology–Hellas,*

*P.O. Box 1527, Vassilika Vouton, 711 10 Heraklion, Crete, Greece*

We present measurements of electrical resistivity, thermal conductivity and thermopower of polycrystalline Sr-doped LaCoO$_3$ with composition La$_{0.95}$Sr$_{0.05}$CoO$_3$. Our data show that the investigated compound exhibits a very respectable room temperature thermoelectric figure of merit value of 0.18. Our results not only show that oxides are promising candidates for thermoelectric cooling applications, but also point towards the need for careful theoretical calculations that will serve as a guide in producing the next generation of thermoelectric materials.

[*]Electronic mail: giapintz@iesl.forth.gr

[a)] *Department of Physics, University of Crete, P.O. Box 2208, 71003 Heraklion, Crete, Greece*

[b)] *Department of Materials Science and Technology, University of Crete, P.O. Box 2208, 710 03 Heraklion, Crete, Greece*

The Seebeck and Peltier effects are non-equilibrium thermodynamic phenomena offering alternative pathways for power generation and refrigeration based solely on solid-state elements [1]. However, a large-scale industrial exploitation of these effects depends on discovering efficient thermoelectric (TE) materials [2]. These materials need to have high Seebeck coefficient, S, low electrical resistivity, $\rho$, and low thermal conductivity, $\kappa$, to exhibit a high figure of merit, $ZT=S^2T/\rho\kappa$, that quantifies the efficiency of a device at a given temperature, T. Modern commercial Peltier devices exhibit a ZT of ~0.9 at room temperature which corresponds to a Carnot efficiency of 10%. It is noteworthy to mention that for solid-state home-refrigeration to be realized thermoelectric devices with a Carnot efficiency of 30% are needed, i.e. thermoelectric materials should be used with ZT$\geq$4. On the other hand, it is encouraging that there is no fundamental thermodynamic obstacle for thermoelectric coolers to reach a Carnot efficiency of up to 100% [2].

Several new bulk materials have been investigated and proposed as potential candidates for TE applications, including skutterudites [5], complex chalcogenides [6], clathrates [7] and half-Heusler alloys [8]. The search for new TE materials has been pursued based on several phenomenological features that have been summarized by F. J. DiSalvo [2]. These features include, among others, a large number of heavy elements in the unit cell of the candidate material, small electronegativity differences between the elements of the compound and the ability to dope the material to an optimal concentration of carriers (~$10^{19}$/cm$^3$). Oxides were largely neglected until Terasaki et al. [9], while working on bronze type oxides as a reference for high temperature oxide superconductors, discovered that the layered cobaltite $NaCo_2O_4$ exhibits a reduced resistivity and a concurrent high thermoelectric power, i.e. it exhibits an unusually high power factor, $S^2/\rho$, for oxides. This initiated the search for efficient TE oxides.

The interest in TE oxides stems from their high thermal stability, excellent oxidation resistance and weak toxicity. To date, the study of oxide materials has mainly focused in the area of high-temperature power generation through the Seebeck effect. Indeed, the room temperature figures of merit for oxide materials are prohibitively low in order for them to be considered for cooling applications, for example $NaCo_2O_4$ has a ZT=0.03 [10], $NaCo_{2-x}Cu_xO_4$ exhibits 0.08 [11] and $Sr_{1-x}La_xTiO_3$ 0.09 [12]. Here, we report that polycrystalline pellets of the pseudo-cubic oxide $La_{0.95}Sr_{0.05}CoO_3$ exhibit a very respectable room temperature ZT value of 0.18. An additional advantage of $La_{0.95}Sr_{0.05}CoO_3$ compared to the aforementioned layered oxides is that it is isotropic (the properties do not depend on crystallographic direction) and thus, there is no need for the cumbersome technique of epitaxial growth in preparing thermoelectric devices.

Ceramic powder samples of $La_{1-x}Sr_xCoO_3$ were prepared by a citrate gel method [13], using very high purity (99.999 %) metal nitrates [($La(NO_3)_3·6H_2O$, $Co(NO_3)_2·6H_2O$, $Sr(NO_3)_2$, Aldrich chemicals] as starting materials. The crystal structure of the synthesized powders was determined by X-ray powder diffraction using monochromatic Cu *Ka₁* radiation (Rigaku RINT 2000). All diffraction patterns showed no evidence for impurity phases within the resolution of the instrument. The stoichiometry of the cations was found to be very close to the nominal one by energy dispersive X-ray spectroscopy (EDX) measurements.

The measurements reported here were performed on bar shaped polycrystalline pellets that were produced at a pressure of 4kbar and sintered at 1000 °C for 5 hours. A final heat treatment at 460 °C in flowing oxygen ensured optimal oxygenation of the pellets that was checked via thermo-gravimetric analysis using a small part of the sintered pellet.

Thermal transport data in the temperature range $80 \leq T \leq 320$ K were obtained in a homemade apparatus employing the steady-state method. Details regarding the apparatus can be found in Ref. 14. The systematic error of $\kappa(T)$ due to the uncertainty in measurements of the sample size and the distance between the thermocouples was estimated to be ~8-10%. Resistivity measurements were collected in a commercial system (Oxford Instruments) with a standard four-probe method.

$LaCoO_3$ is a p-type semiconducting material, which exhibits a rhombohedrally distorted perovskite structure [15]. In the past this class of cobaltite compounds have been extensively studied because they exhibit quite peculiar magnetic and transport properties that have been associated with a thermally activated spin state transition from a low spin (S=0) ground state configuration to an intermediate (S=1) and to a high (S=2) spin state with increasing temperature. Actually, this unusual spin state configuration dependence on several external factors has been proposed [16] as the main reason for the enhanced Seebeck coefficient values found in rare-earth cobaltites (see Fig. 2 where filled squares represent the values of $LaCoO_3$) and alkaline earth substituted cobaltites [17-21] even at elevated temperatures. Substituting $Sr^{2+}$ for $La^{3+}$ brings about remarkable changes in the system, the most prominent being the gradual evolution of a metallic almost long-range ferromagnetic state along with a gradual reduction of the rhombohedral distortion [15].

La$_{0.95}$Sr$_{0.05}$CoO$_3$ is a semiconducting compound. Figure 1 illustrates the temperature dependence of the resistivity of a polycrystalline pellet. The resistivity of the compound follows the Arrhenius law for thermally activated conduction, $\rho(T) \sim \exp\{-E_g/k_BT\}$, where $E_g$ is the activation energy barrier and $k_B$ is the Boltzmann constant. Plotting the logarithm of the resistivity, ln$\rho$, as a function of the inverse temperature, 1/T, one obtains a straight line of slope -$E_g/k_B$. In our case $E_g$ has been calculated to be 36 meV, which is an order of magnitude lower than that calculated for the parent compound LaCoO$_3$ (250 meV) [22]. This observation indicates that replacing Sr for La cations in the LaCoO$_3$ lattice induces considerable electronic changes in addition to modifying the carrier concentration of the parent compound. This is further supported by a close examination of the temperature dependence of the Seebeck coefficient.

Figure 2 presents the Seebeck coefficient, S, of La$_{0.95}$Sr$_{0.05}$CoO$_3$ (open circles) as a function of temperature. We observe that the investigated compound exhibits an exceptionally large room-temperature thermoelectric power response [S(300K) ~ 710 µV/K] a fact that makes it particularly attractive for thermoelectric applications. Note that thermoelectric power measurements of Ohtani et al. [17] on polycrystalline samples as well as of Kobayashi et al [21] on single crystals of La$_{0.95}$Sr$_{0.05}$CoO$_3$ agree with the ones reported here.

The Seebeck coefficient determines the average energy, with respect to the Fermi level $E_F$, which is transported by charge carriers under the influence of a thermal gradient. Therefore, it can be considered as a direct probe of the changes occurring at the Fermi level. In the general case, S is expressed as

$$S = (\mu - <E>) / (|e| \times T) \quad (1)$$

where μ is the chemical potential and <E> is an average energy of the carriers weighted by their contribution to the conductivity. In the case of a band semiconductor in which the relaxation energy of the carriers, E, is described by a relaxation time, $\tau \sim E^r$, the Seebeck coefficient takes the form:

$$S = \frac{k_B}{e}\left[\frac{5}{2} + r + \frac{E_F - E_V}{k_B T}\right] \quad (2)$$

where $E_V$ is the valence band energy. Therefore, we propose that Sr substitution for La induces several changes in the Fermi level of the parent compound that may stem from considerable changes in the band structure. To clarify this issue there is a need for detailed band structure calculations.

Figure 3 shows the thermal conductivity, κ, of $La_{0.95}Sr_{0.05}CoO_3$ as a function of temperature. κ rises from 0.016 W/cm-K at 80 K to 0.030 W/cm-K at 300 K. This unexpected temperature dependence of κ has also been observed for other oxides [18]. Occasionally it has been attributed to radiation losses in which case the reported κ values constitute an upper bound for our investigated samples. However, there are indications that this behavior may be intrinsic in which case it is worth further investigations. Using the measured values of S, ρ and κ we have calculated the dimensionless figure of merit ZT (depicted in Fig. 3). The room temperature ZT of 0.18 is the highest reported for an oxide to date. It is noteworthy to point out that ZT exhibits a monotonic linear increase from ~140 K to room temperature and thus, we expect it to reach its maximum quite above room temperature. It is therefore important that this compound be investigated at higher temperatures since there are strong indications it may be a good candidate for power generation applications as well.

The observed ZT value is directly comparable to current bulk state-of-the-art thermoelectric materials such as bismuth telluride, which exhibits a room temperature figure of merit of ~0.9. This result is significant since it suggests that room temperature thermoelectric oxides may be realized. It should be emphasized that the reported values were extracted from polycrystalline pellets and thus they are not optimal since several effects related to such type of samples (e.g. grain boundary resistance) may significantly alter the values obtained from single crystals. Calculations using single crystal resistivity values [21] show that ZT will increase and may reach values as high as 0.5 depending on how the thermal conductivity will increase in a single crystal. In addition, the present work shows that a large figure of merit may be possible in materials consisting of relatively light atoms, such as Co and O, and exhibiting high electronegativity differences between the constituent elements (i.e., reduced covalency of the bonds) in striking contrast to the common sense that a large number of heavy elements per unit cell forming highly covalent bonds will likely constitute the next generation thermoelectric material. Therefore, future approaches regarding the discovery of efficient TE materials should heavily rely on the careful examination of the electronic band structure.

In summary, we present transport data $La_{0.95}Sr_{0.05}CoO_3$ polycrystalline samples, which indicate that this material may serve as a potential room-temperature p-type thermoelectric oxide for cooling applications. Our results point towards the need for careful theoretical calculations that will serve as a guide in producing the next generation of thermoelectric materials.

**Figure Captions**

Fig. 1 Temperature dependence of the resistivity for the polycrystalline bar-shaped sample of $La_{0.95}Sr_{0.05}CoO_3$. Note that it exhibits a room temperature resistivity of 27 m$\Omega$cm.

Fig. 2 Thermoelectric power response of the polycrystalline bar-shaped sample of $La_{0.95}Sr_{0.05}CoO_3$ as a function of temperature. Note the qualitative and quantitative differences induced by 5 % La substitution by Sr (open circles) in contrast to the parent $LaCoO_3$ oxide (filled squares).

Fig. 3 Thermal conductivity and dimensionless thermoelectric figure of merit as a function of temperature for the polycrystalline bar-shaped sample of $La_{0.95}Sr_{0.05}CoO_3$. The observed ZT at room temperature is the highest reported for an oxide material.

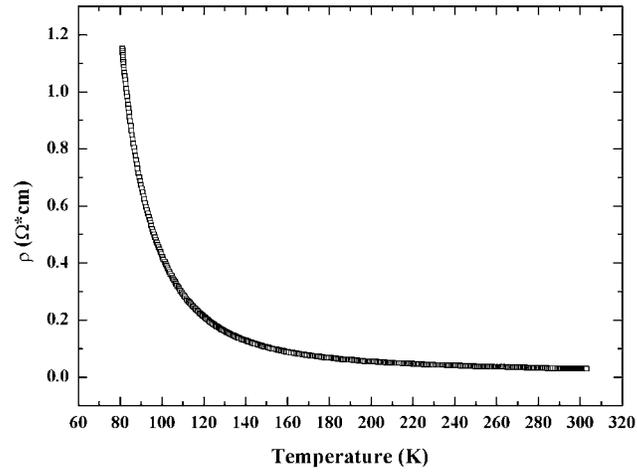

**Figure 1**: J. Androulakis et al. Applied Physics Letters

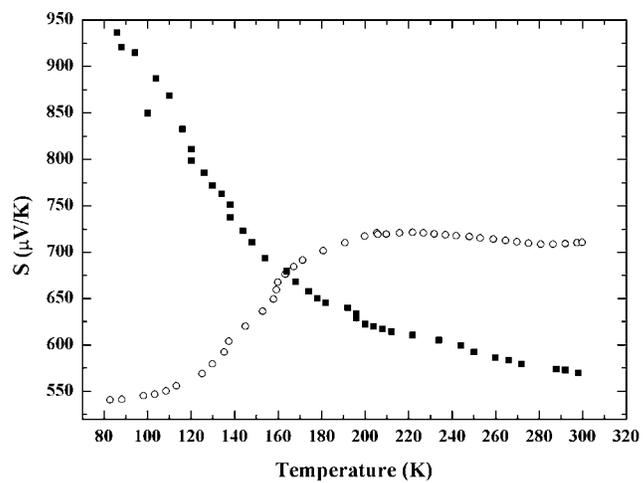

**Figure 2**: J. Androulakis et al. Applied Physics Letters

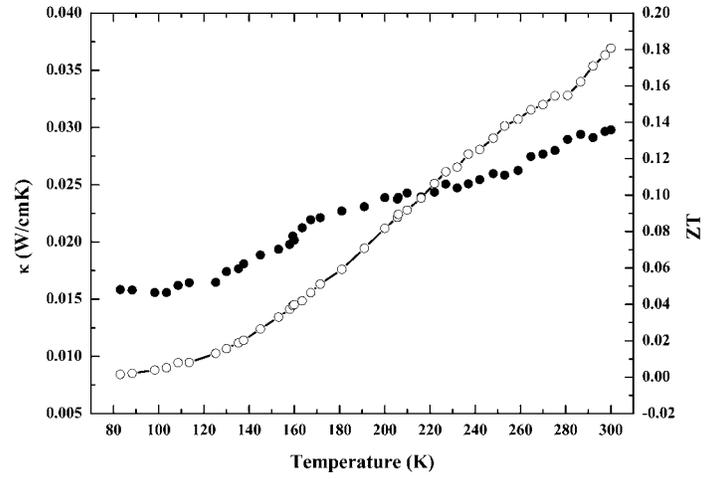

**Figure 3** J. Androulakis et al. Applied Physics Letters